# On the origin of the light yield enhancement in polymeric composite scintillators loaded with dense nanoparticles.

Irene Villa[*,1], Angelo Monguzzi[*,1,2], Roberto Lorenzi[1], Matteo Orfano[1], Vladimir Babin[3], František Hájek[3], Karla Kuldová[3], Romana Kučerková[3], Alena Beitlerová[3], Ilaria Mattei[4], Hana Buresova[5], Radek Pjatkan[5], Václav Čuba[6], Lenka Prouzová Procházková[3,6], and Martin Nikl[3]

[1]*Dipartimento di Scienza Dei Materiali, Università Degli Studi Milano-Bicocca, 20125 Milano, Italy*

[2]*NANOMIB, Center for Biomedical Nanomedicine, University of Milano-Bicocca, P.zza Ateneo Nuovo 1, 20126 Milan, Italy*

[3]*FZU─Institute of Physics of the Czech Academy of Sciences, Cukrovarnická 10/112, 16 200 Prague, Czech Republic*

[4]*INFN Sezione di Milano, via G. Celoria 16, 20133, Milan, Italy*

[5]*NUVIA a.s, Trojanova 117, 278 01 Kralupy nad Vltavou, Czech Republic*

[6]*Faculty of Nuclear Sciences and Physical Engineering CTU in Prague, Czech Republic*

Corresponding authors:  irene.villa@unimib.it , angelo.monguzzi@unimib.it

**Abstract**

Fast emitting polymeric scintillators are requested in advanced applications where high-speed detectors with large signal-to-noise ratio are needed. However, their low density implies a weak stopping power of high energy radiations, thus a limited light output and sensitivity. To enhance their performances, polymeric scintillators can be loaded with dense nanoparticles (NPs). We investigate the properties of a series of polymeric scintillators by means of photoluminescence and scintillation spectroscopy, comparing standard scintillators with a composite system loaded with dense hafnium dioxide ($HfO_2$) NPs. The nanocomposite shows a scintillation yield enhancement of +100% with unchanged time response. We provide for the first time an interpretation of this effect, pointing out the local effect of NPs in the generation of emissive states upon interaction with the ionizing radiation. The obtained results indicate that coupling of fast conjugated emitters with optically inert dense NPs could allow to surpass the actual limits of pure polymeric scintillators.

Fast and highly emissive plastic scintillators, usually made of a polymeric scintillating matrix that host a fluorescent dye, are requested for many advanced applications where high signal-to-noise ratio is required in a short time window. For example, to detect high rate events avoiding pile up in high energy physics experiments at the energy and intensity frontiers to face the challenges of unprecedented event rate and severe radiation environment,[1,2,3] or to quickly acquire high quality image at low dose in medical applications as in the time-of-flight positron emission tomography (TOF-PET) imaging technique, where tents of picoseconds time resolutions are desired.[1,4-6] Unfortunately, their low density results a low stopping power of high energy radiation and their scintillation light yield $\phi_{LY}$, defined as the ratio between the number of emitted photons and the energy deposited in the system, is lower than the one of best inorganic scintillators (Supporting Information, Table 1). This results in a detrimental reduction of the emitted light output and consequently of the detector sensitivity, especially in the case where small sensors are required.[7]

A common strategy to improve their $\phi_{LY}$ is the use of wavelength shifting dyes, which harvest more efficiently the energy deposited in the host matrix by the ionizing radiation and populate the final emitters through several energy transfer processes.[8] The emitters can be therefore used at low concentration, releasing from self-absorption problems with beneficial consequences on the light output. However, due to the multiple energy transfer steps involved, this strategy could result in a delay of the scintillation time response detrimental for fast applications. On the other hand, the use of hybrid organic/inorganic compounds [9,10,11,12,13] or the loading of polymeric hosts with optically inert dense nanoparticles (NPs) are alternative approaches recently proposed to enhance the stopping power of liquid and polymeric conjugated scintillators.[14-19,20-23] In principle, the presence of dense NPs including high atomic number $Z$ element would affect indeed both the Compton and the photoelectric interactions,[24] because the interaction probability with high energy photons, such as X-rays and γ-rays, quickly increases with the effective material electronic density and the effective $Z$ value. Rather than to create luminescence centers, the role of these inorganic fillers is therefore to enhance the efficacy of the transformation of the absorbed energy into electronic excitations, which will then be transferred through non-radiative mechanisms to the final emitter.[25,26] In particular, NPs-loaded composites scintillators show a composition-dependent behavior,[27,28,29] and trade-off conditions that maximize the $\phi_{LY}$ can be found.[30,31] However, a clear explanation of the effect that correlates the presence of NPs with the scintillation performance is still under debate.

In order to point out the mechanism behind the $\phi_{LY}$ enhancement observed in composite systems, we investigate here the properties of a model polymeric scintillator based on atactic polystyrene (PS) and doped with the scintillating dye 1,4-bis(5-phenyloxazol-2-yl) benzene (POPOP).[32] The system has been modified firstly in a traditional way by adding the standard energy harvesting scintillating dye p-terphenyl (TP). In the second case, we loaded it with high density hafnium dioxide ($HfO_2$) nanoparticles (NP, $\rho$ = 9.68 g cm$^{-3}$) - already employed to enhanced the sensitivity of liquid and hybrid scintillators as well as radiosensitizers in cancer treatment - [15,33-36] to obtain a composite material. The scintillation mechanism has been modelled considering the yield of the energy transfer processes involved as a function of the dyes concentration, and comparatively studied by means of steady state and time resolved photoluminescence and scintillation spectroscopy. The scintillation enhancement of +100% achieved in the composite material, under both soft X-rays and γ-rays irradiation (see Supporting Information, SI, section 6), is comparable to the one obtained using the intermediate dye TP as energy harvester/transporter, but leaving unchanged the scintillator time response. Notably, this happens despite the average density of the material is substantially unchanged upon a low NPs loading. We provide a possible interpretation of this sensitization effect, based on the local effect of the presence of NPs in the radiation-matter interaction dynamics, which confirms that a finely controlled loading with optically inert dense NPs could be effective to control the properties of a new generation of fast polymeric scintillator and metascintillators.

Figure 1a depicts the absorption and photoluminescence spectra of the PS host matrix, of the POPOP and the TP scintillating conjugated dyes.[8] The sample thickness is 1 mm (SI, section 2). The reference PS:POPOP scintillator contains 0.05% of weight fraction (%wt) of POPOP. The second full organic sample PS:TP:POPOP contains 0.6%wt of TP and 0.05%wt of POPOP. The nanocomposite sample PS:NP:POPOP contains 1%wt of NPs and 0.05%wt of POPOP. The NPs were prepared via photo-induced synthesis, based on the reaction of dissolved salts with products of water photolysis (SI, section 2).[37] Powder X-rays diffraction and Raman experiments indicates that upon annealing at 1000 °C the NPs show a pure monoclinic crystalline phase with space group $P2_1/c$ (Supporting Figs. S1-S3).[38-40]

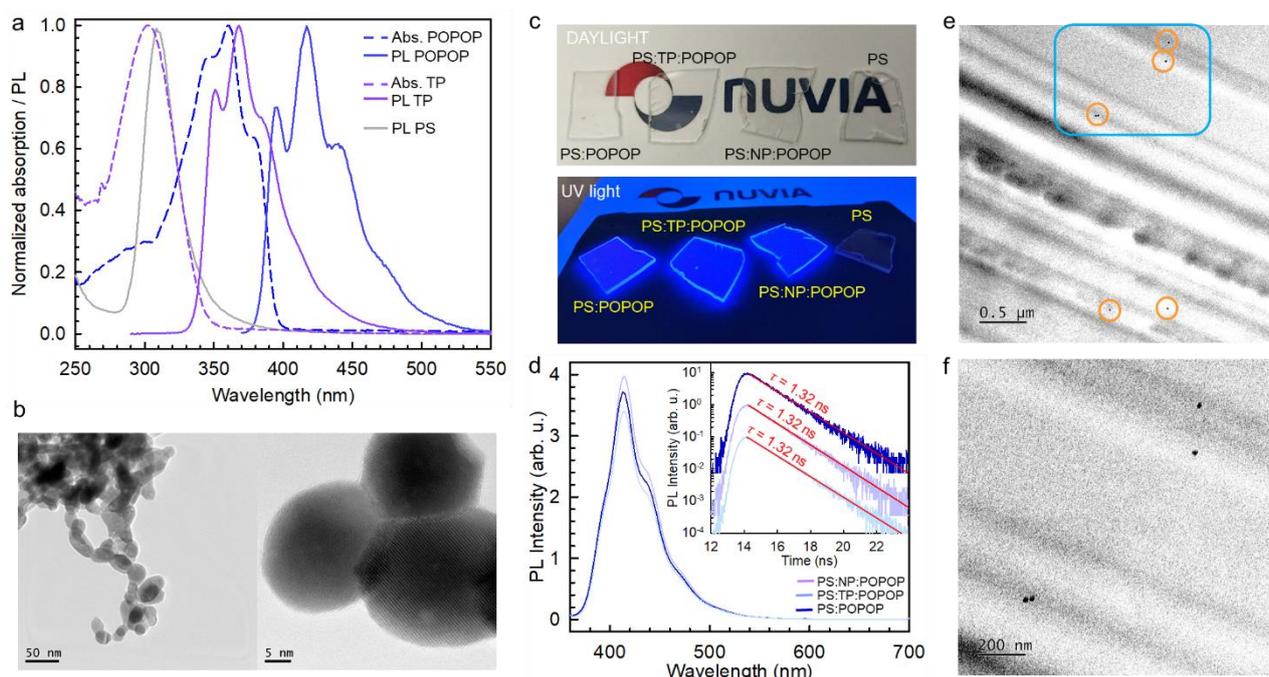

**Figure 1.** (a) Absorption and photoluminescence (PL) spectra of PS, POPOP and TP in tetrahydrofuran diluted solution ($10^{-6}$ M). (b) High resolution transmission electron spectroscopy (TEM) images of the synthesized HfO2 nanoparticles (NPs). (c) Digital pictures of the investigated plastic scintillators under daylight and UV excitation at 365 nm. The NUVIA logo is reproduced under permission, copyright@2024 Nuvia a.s. (d) PL spectra of the scintillator series under cw excitation at 340 nm. The inset shows the PL intensity decay in time recorded at 420 nm under pulsed excitation at 340 nm (pulse width 500 ps). Solid lines are the fit of data with a single exponential decay function of characteristic decay time t. (e) High resolution TEM image of sample PS:NP:POPOP. The circles mark the loaded NPs. (f) Magnification of the blue larked area of panel e.

The NPs show a negligible photoluminescence and radioluminescence (Fig.S5). Importantly, their size of 25 nm (Fig.1b and Fig.S1) is small enough to allow all the energy deposited by primary or secondary interactions with high energy radiations within the NPs to escape from the particle in the surrounding PS where they are dispersed (Fig. 1e,f).[41] Under UV excitation all samples, except for the PS scintillator, emit blue light (Fig.1c). Figure 1d shows the photoluminescence spectra of the samples series investigated under direct excitation of POPOP molecules at 340 nm, where TP and PS are transparent. Given their low amount ($1.6\times10^{-7}$ M) and their wide bandgap, the NPs absorption is negligible at this wavelength (Fig.S6). The samples photoluminescence intensity is constant within the experimental uncertainty of ±10%. The POPOP emission lifetime $\tau = 1.32$ ns is identical in all samples (Fig.1d, inset) and matches the one in diluted solution.[31] These findings demonstrate that upon direct optical excitation, the POPOP photoluminescence properties are preserved in the PS host, including its high photoluminescence quantum yield $\phi_{pl} = 0.93$.[42-44]

The photoluminescence efficiency is only one of the parameters that set the scintillation efficiency. Specifically, in a multicomponent material designed for sensitized scintillation, the scintillation emission intensity $I_{scint}$ can be expressed as a function of the sensitizer and emitter concentration, $C_S$ and $C_E$, respectively, by

$$I_{scint}(C_S, C_E) \propto \kappa \phi_{LY}(C_S, C_E) = \kappa[N\chi(\varepsilon\phi_S + \phi_E)\phi_{pl}], \qquad \text{Eq. 1}$$

where $N$ is the number of charge carrier pairs generated in the material upon interaction with the ionizing radiation and $\kappa$ is the instrumental detection efficiency. All the other parameters are efficiency indicators that varies from 0 to 1. The factor $\chi$ measures the conversion efficiency from plasma states of free charge carriers to optical photons, while the parameters $\phi_S$ and $\phi_E$ are the efficiencies of the energy harvesting ability of the sensitizer and emitter, respectively. Lastly, $\varepsilon$ marks the efficiency of the energy transfer from excited sensitizers to emitters.[27] As shown in Fig.2a, in the PS matrix the free charge carriers recombine to create an excimer state $PS^{D*}$ with emission peaked at 320 nm.[45, 46] The POPOP luminescence is activated by homo-molecular diffusion-mediated Förster energy transfer from $PS^{D*}$, in agreement with the concentration of conjugated rings in the PS matrix (~10 M) that makes them by far the most effective charge capture center. Given the long $PS^{D*}$ lifetime $\tau_{PS}^{D*} = 7$ ns (Fig.S7), the diffusivity of PS excitons and the concentration of POPOP,[47] the PS→POPOP energy transfer process occurs in the rapid diffusion limit. The corresponding total energy transfer rate $k_{Fs}^{PS\rightarrow POPOP}$ is then given by

$$k_{Fs}^{PS\rightarrow POPOP} = \frac{4\pi(R_{Fs}^{PS\rightarrow POPOP})^6 C_{POPOP}}{3\tau_{PS}^{D*}a^3}, \qquad \text{Eq. 2}$$

where $a = 1.1$ nm is the minimum center-to-center distance between the PS conjugated ring and the POPOP molecule.[48, 49] As shown in Fig.2b, the PS→ POPOP energy transfer efficiency calculated as

$$\phi_{ET}^{PS\rightarrow POPOP} = \frac{k_{Fs}^{PS\rightarrow POPOP}}{k_{PS} + k_{Fs}^{PS\rightarrow POPOP}} \qquad \text{Eq. 3}$$

reproduces the experimental data obtained as a function of the POPOP concentration (Fig.S8) by considering a Förster radius $R_{Fs}^{PS\rightarrow POPOP} = 3.2$ nm, in agreement with the expected value of 2.8 nm calculated from the PS and POPOP emission and absorption properties (Fig.1a and SI, section 4).[50] This demonstrates that the excitation of POPOP by direct recombination of diffusing charges on the dye is a negligible pathway. This conclusion is further supported by looking at Fig. 2b showing the direct charge capture yield $\phi_{e-h}^{\square}$. This efficiency, calculated considering a typical charge capture radius of $R_{e-h}^{\square} = 15$ nm in a polymeric host,[51,52] should be close to unit at POPOP concentration of $10^{-6}$ M, far below the experimental condition where all the investigated samples contains $1.4\times10^{-3}$ M of POPOP. At this concentration, the experiemental $\phi_{ET}^{PS\rightarrow POPOP}$ results 0.95, with a total transfer rate of $k_{Fs}^{PS\rightarrow POPOP} = 2.8$ GHz derived from Eq.3. The system measured $\phi_{LY}$ under soft X-rays is as high as 1340±250 ph MeV$^{-1}$ (*vide infra* Fig.4a and SI).

The classical strategy to improve $\phi_{LY}$ for polymeric scintillators is to use a second scintillating dye as energy collector and transporter,[53] to better harvest the deposited energy from the matrix while avoiding self-absorption effects. We opted here for the TP, whose absorption spectrum is resonant with the $PS^{D*}$ emission at 320 nm (Fig.1b) resulting a PS→TP Förster radius as large as $R_{Fs}^{PS\rightarrow TP} = 16.2$ nm (SI, section 4). The TP concentration in the sample is $2.7\times10^{-2}$ M, which implies that that the energy stored in the PS$^{D*}$ states is completely and quickly transfered to TPs, with a rate $k_{Fs}^{PS\rightarrow TP} \gg k_{Fs}^{PS\rightarrow POPOP}$ in the THz range, instead of being directly trasferred to POPOPs (Fig.2b). The subsequent excitation of POPOP molecules occurs by re-asborption of the TP luminescence (Fig.S5), given the poor efficiency of the TP→POPOP non-radiative transfer (Fig.1c).

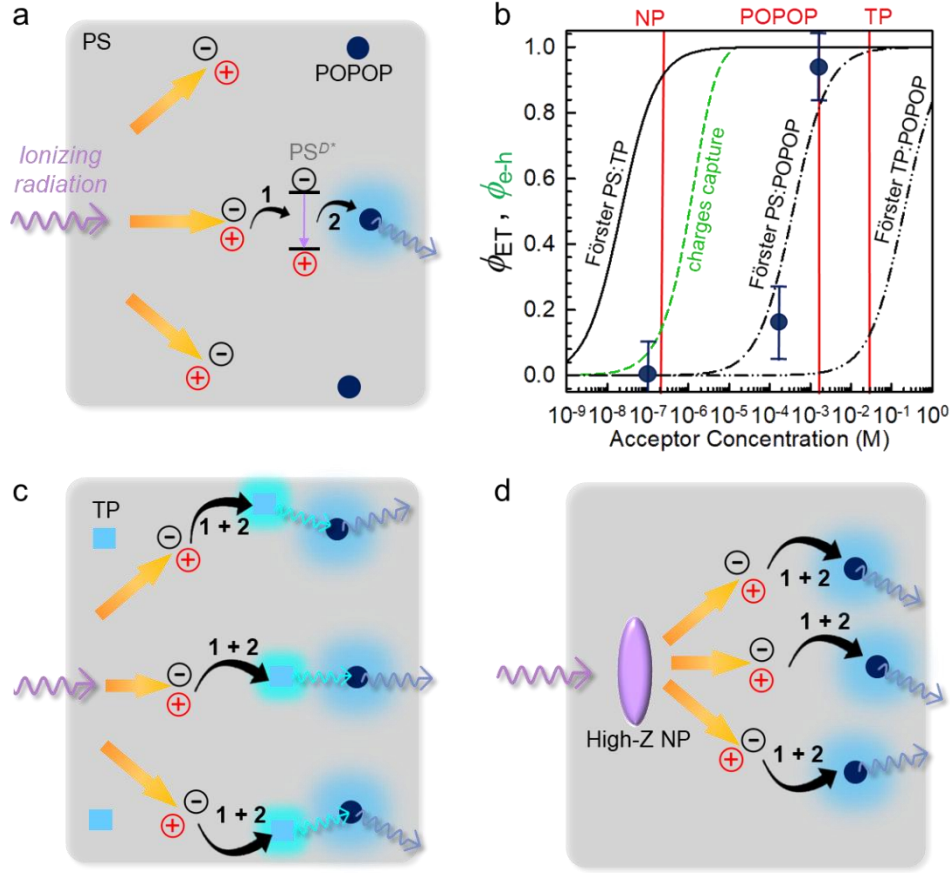

**Figure 2.** (a) Sketch of the scintillation mechanism in standard polymeric scintillator composed by polystyrene (PS) as host for the scintillating dye POPOP ($2.4\times10^{-2}$ M). The diffusing charges generated upon interaction with the ionizing radiation recombine to form PS excimers, PS$^{D*}$ (step **1**). The POPOP dyes are subsequently excited by non-radiative Förster energy transfer from the PS$^{D*}$ states (step **2**). (b) Black lines show the calculated Forster energy transfer efficiency for different energy donor/acceptor pairs present in the sample investigated (PS:POPOP, PS:TP, TP:POPOP) as a function of the acceptor concentration. Black dots mark the efficiency of the PS→POPOP energy transfer measured a function of the POPOP concentration with has been fitted with Eq.3 using an interaction radius $R_{Fs}^{PS\rightarrow POPOP}$ = 3.2 nm. The green line shows the calculated charge capture yield as a function of the hafnia NPs concentration considering a charge capture radius of $R_{e-h}$ = 15 nm. The red vertical lines mark the POPOP, TP and NPs concentrations employed in the investigated samples for a comparative analysis. (c) Sketch of the scintillation mechanism in a polymeric scintillator made of PS, TP ($2.7\times10^{-2}$ M), and POPOP ($1.4\times10^{-3}$ M). After steps **1** and **2** towards TP molecules, the POPOP dye is excited by re-absorption of the TP emission. (d) Sketch of the scintillation mechanism in a polymeric scintillator made of PS, POPOP ($1.4\times10^{-3}$ M) and NPs ($1.6\times10^{-7}$ M). The POPOP dyes are excited by non-radiative Förster energy transfer from the PS$^{D*}$ through steps **1** and **2** and by exploiting an improved interaction with the ionizing radiation mediated by NPs.

As alternative to TP, we loaded the PS:POPOP scintillator with dense HfO$_2$ NPs in order to promote the interaction with high energy photons. The NPs amount is as low as $1.6\times10^{-7}$ M in order to preserve the material optical quality (Fig.S5), thus the average density of the PS:NP:POPOP scintillator matches the one of the PS:POPOP and PS:TP:POPOP samples. Nevertheless, the system shows an improved $\phi_{LY}$ = 2460±300 ph MeV$^{-1}$ under soft X-rays, twice the reference system (Fig.4a). Considering that i) the low concentration of NPs limits the yield of diffusing charges capture to < 0.1 (Fig.2b) and ii) their poor emission properties prevent energy transfer to POPOP molecules, we ascribe the observed $\phi_{LY}$ enhancement to a peculiar sensitization effect.

We investigate more in detail the properties of the scintillators to shed light on this point and highlight possible different activation mechanisms in the nanocomposite with respect to the TP-loaded system. By means of a Monte Carlo simulations of the radiation-matter interaction, we evaluate how the energy is released in the samples (SI, section 1).[54, 55] Figure 3 shows the simulated distribution of the energy loss events. The total fraction of the energy $E_{tot}$ released by 7 keV photons is ~96% in all the three cases, thus enabling quantitative comparisons. Notably, only with soft X-rays we can be sure of the amount of energy deposited in these thin plastic scintillators, thus corroborating the reliability of the following discussion. Conversely, the correct estimation of the energy deposition in a thin polymeric scintillator is quite difficult with higher energy photons, implying an intrinsic uncertainty that would affect the quantitative absolute comparisons. Interestingly, the events distribution for the full organic systems PS:TP:POPOP and PS:POPOP is identical, with the majority of the events matching the photoelectric peak at 7 keV. On the other hand, despite their low concentration, the presence of NPs generates a broader distribution of events at lower energies, down to 5 keV. The simulations suggest therefore a different interaction of high energy photons with dense and high-Z NPs, which induces a different distribution of the total energy in the nanocomposite volume.

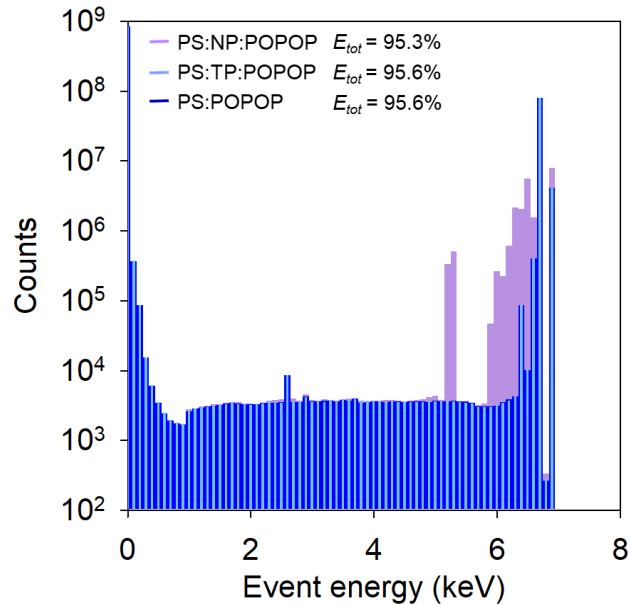

**Figure 3.** Energy loss events probability calculated by a Monte Carlo simulation for the three scintillators investigated (PS:POPOP, PS:TP:POPOP, PS:NP:POPOP) for 7 keV photons. $E_{tot}$ is the fraction of energy released in the material with respect to the intensity of the source employed in the simulation.

Equation 4 shows how the relative scintillation yield $\eta$ in different materials can be expressed as a function of the parameters that define the efficiency of the energy-to-photon conversion in a multicomponent scintillator (Eq. 1) as:

$$\eta = \frac{\phi_{LY}^X(PS:X:POPOP)}{\phi_{LY}(PS:POPOP)} = \frac{\chi^X N^X (\varepsilon \phi_X + \phi_{POPOP})}{\chi N \phi_{POPOP}} \frac{\phi_{pl}^X}{\phi_{pl}} \quad .\qquad \text{Eq. 4}$$

Here, the PS:POPOP scintillator is the reference, while $X$ marks the comparative system. In the PS:TP:POPOP scintillator, dealing with a fully organic system where diffusing charge recombine mainly forming the PS excimers as in the reference, we can consider $\chi^{TP} = \chi$. The parameter $\phi_{POPOP}$ is zero, because the energy is harvested by TP molecules with $\phi_{TP} = 1$ (Fig.2b) and subsequently transferred radiatively to POPOP with yield $\varepsilon = 1$ (Fig.S6). Conversely, at the denominator we have $\phi_{POPOP} = 1$, thus

$$\eta^{TP} = \frac{\phi_{scint}^{TP}(PS:TP:POPOP)}{\phi_{scint}(PS:POPOP)} = \frac{N^{TP}}{N}\frac{\phi_{pl}^{TP}}{\phi_{pl}} = \frac{\phi_{pl}^{TP}}{\phi_{pl}}. \qquad \text{Eq. 5}$$

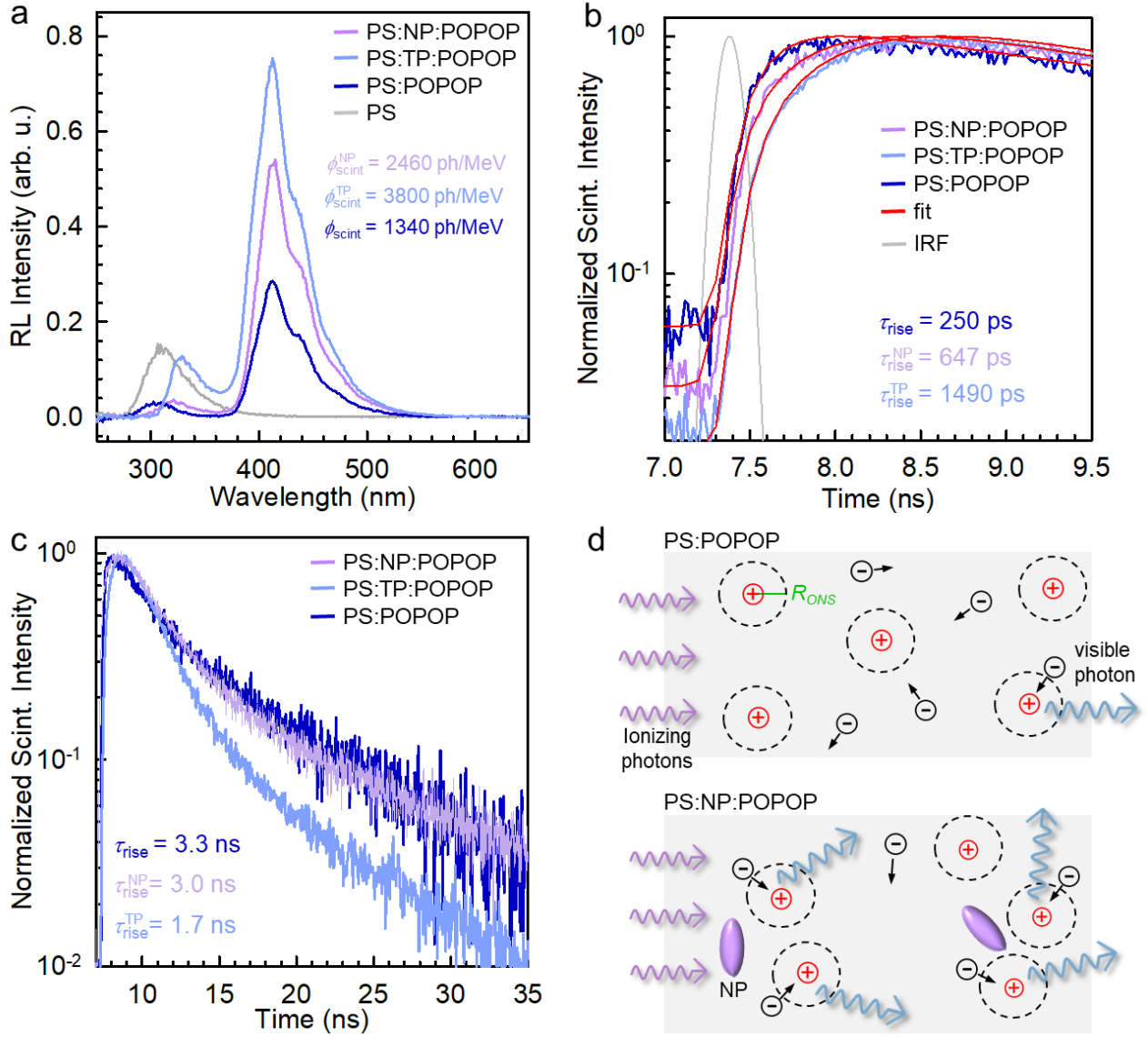

**Figure 4.** (a) Radioluminescence (RL) spectra of PS:POPOP, PS:TP:POPOP, PS:NP:POPOP scintillators vs. pure polystyrene (PS) under excitation with soft X-rays (7 keV). (b) Rise time of the scintillation emission measured for the scintillator series under ultrafast excitation with soft X-rays (14.5 keV, pulse width 120 ps). Solid lines are the fit of the data with a multi-exponential decay function convoluted with the instrumental response function (IRF) employed to estimate the signal rise times. (c) Scintillation pulse recorded at 420 nm recorded under ultrafast excitation with soft X-rays. (d) Sketch of the diffusing free charges recombination mechanism in absence (top) and in presence (bottom) of dense nanoparticles (NP). Even if the total number of free charges is the same, in the surrounding of NPs the density of charges is higher, thus the probability that they diffuse and thermalize to an intermolecular distance shorter than the recombination Onsager radius $R_{ONS}$ to form emissive molecular excitons is higher.

From the data in Fig. 4a we know that $\eta^{TP} = 2.8$, and the full-organic composition of the materials means that $N^{TP} = N$ (Fig.3). Thus, Eq. 5 suggests that the improved $\phi_{LY}$ of the PS:TP:POPOP sample is due to a better emission efficiency. The data in Fig.4 support this view. Figures 4b and c show the scintillation kinetics recorded at 420 nm for the samples investigated under pulsed soft X-rays. The PS:POPOP emission intensity rises up in $\tau_{rise}$ = 250 ps (Fig.4b). This means that the POPOP emissive state is populated with a rate $k_{rise} = (\tau_{rise})^{-1} = 4.0$ GHz, in agreement with the experimentally-derived predicted PS→POPOP Förster transfer rate and therefore supporting the scintillation mechanism proposed (Fig.2a). On the other hand, the scintillation emission intensity decays as multi-exponential function with a characteristic decay time of $\tau_{scint}$= 3.3 ns, calculated as the time at which the time-integrated emission intensity is reduced to a factor 1/$e$ (Fig.4c). This value is more than twice than the intrinsic POPOP emission lifetime, confirming the behavior previously observed for PS-embedded POPOP molecules upon X-rays.[31] We ascribe this behavior to a possible exciplex formation between the excited PS and POPOP molecules,[56] but further studies are still ongoing. Nevertheless, usually exciplexes are not efficient emitters and their photoluminescence efficiency is lower with respect to the single molecule one, so it is realistic to assume that the system photoluminescence quantum yield $\phi_{pl}$ is lower than 0.93 under high energy excitation. On the other hand, the PS:TP:POPOP scintillation shows a slow rise time $\tau_{rise}^{TP}$ = 1.5 ns, matching the TP emission lifetime of 1.2 ns (Fig.S7), thus demonstrating the activation of the POPOP emission by re-absorption of TP luminescence (Fig.2c). This is confirmed by the scintillation decay time value, which is as fast as $\tau_{scint}^{TP}$= 1.7 ns. This value is very similar to the one of the optically-activated POPOP emission (Fig.1d, inset), i.e. the condition where the POPOP shows its best luminescence properties. Thus, we can assume here $\phi_{pl}^{TP}$= 0.93. These findings demonstrate that i) in the PS:TP:POPOP system the scintillation light is activated mostly by the TP→POPOP radiative energy transfer, and that ii) $\phi_{scint}^{TP} > \phi_{scint}$ because $\phi_{pl}^{TP} > \phi_{pl}$.

A different picture can be envisaged for the nanocomposite PS:NP:POPOP. The scintillation intensity decays with a kinetic similar to that one of the reference with a similar decay lifetime $\tau_{scint}^{NP}$= 3.0 ns (Fig.4c). The scintillator emission properties are therefore basically the same with and without NPs, thus $\phi_{pl}^{NP} = \phi_{pl}$. Given the NPs low amount and their low photoluminescence efficiency, both $\phi_{NP}$ and $\varepsilon$ in Eq. 4 equal zero. Therefore, with NPs we have

$$\eta^{NP} = \frac{\phi_{LY}^{NP}(PS:NP:POPOP)}{\phi_{LY}(PS:POPOP)} = \left(\frac{\chi^{NP}}{\chi}\right)\left(\frac{N^{NP}}{N}\right). \qquad \text{Eq. 6}$$

From the data in Fig.4a we have $\eta^{NP}$= 1.8. Eq. 6 suggests therefore that in the composite the improved $\phi_{LY}$ is due to the generation of a larger number of free charge carriers or to their more efficient conversion to emissive excitons. It is worth noting that the scintillation signal rises up still in the sub-nanosecond time scale with $\tau_{rise}^{NP}$ = 647 ps, but slightly slower that the reference systems. This suggests a more complex process of generation of emissive states, however, as demonstrated by simulations (Fig.3), we do not expect an extraordinary interaction with the ionizing radiation in presence of NPs, so again $N^{NP} = N$. Therefore, the obtained results propose that the sensitization of composite scintillation could be due to a local effect that results a $\chi^{NP} > \chi$.

Figure 4d shows a sketch of a simplified physical modelling of the scintillation mechanism in nanocomposites that can support this picture. The key point is to consider the punctual interaction of high energy photons or secondary electrons with a dense NP.[57] On average, the total energy released in the material is the same with or without NPs (Fig.3), but the local distribution can be significantly due to the nine time larger density of NPs with respect to PS and to the presence of high-Z elements such as hafnium. After interaction, around the NPs we have therefore an initial larger density of diffusing free charges with respect to a pure polymeric system. Thus, in agreement with the Onsager theory,[58,57] the cumulative probability that diffusing charges recombine to form molecular excitons is larger because they thermalize most likely at intermolecular distances shorter than the recombination capture radius, i.e. the Onsager radius $R_{ONS}$ (Fig.4d). We can explicit the formal relationship between $\phi_{LY}$ and its linear electron energy deposit $dE/dx$ as

$$\phi_{LY} \propto \int_0^{E_{tot}} \frac{1-N\exp\left(-\frac{dE/dx}{(dE/dx)_{ONS}}\right)}{1+kB(dE/dx)} dE. \qquad \text{Eq.7}$$

Here $E_{tot}$ is the total energy deposited in the scintillator, constant in our experiments. The parameter $kB = (dE/dx)^{-1}_{Birks}$ is the empirical Birks factor that accounts for the bimolecular quenching processes appearing when high densities of charge carriers are produced. Specifically, the Birks term indicates the $dE/dx$ value for which the average electron-electron spacing is shorter than their diffusion length during thermalization, thus activating a bimolecular collisional quenching mechanism. This term points out the origin of the sub-linear behavior of $\phi_{LY}$ vs. energy of the ionizing beam when >100 keV photons are employed with dense materials.[59] In polymeric scintillators, it can be considered negligible when < 20 keV photons are used.[58]

On the other hand, the Onsager term $(dE/dx)_{ONS}$ in the numerator dominates at low $E_{tot}$ values as in our experiments. It marks the $dE/dx$ value at which the average electron-hole spacing equals $R_{ONS}$. According to Eq.1, a too small $dE/dx$ reduces the fraction $\chi$ of charge carrier pairs converted to emissive excitons with respect the initial value $N$, because of a non-efficient electron-hole recombination. This term could in principle describe the mechanism of the non-linear response of plastic scintillators to low energy photons beams.[60] In the nanocomposite, the local $dE/dx$ value abruptly rises up where the interaction with NPs happens. Thus, even if the total number of generated free charge carriers is the same, in this case they are created in a smaller volume in the NPs surroundings (Fig.4d). In other words, we have locally a larger density of diffusing charges, which implies a final average electron-hole spacing more probably below $R_{ONS}$ after thermalization. This condition can therefore mitigate the effect of the Onsager loss inducing the observed enhancement of $\phi_{LY}$ in the nanocomposite, but without affecting the emission decay time. This picture, even if not supported by a detailed computational simulation tools, could tentatively describe the behavior of scintillating nanocomposites justifying both the efficiency increment and the preservation of the scintillation pulse kinetics.

In conclusion, the spectroscopic investigation performed on a series of polymeric composite scintillators enables to shed light on the effect of their composition on the scintillation efficiency and kinetics. Several strategies can be pursued to enhance the scintillation yield of a polymeric material, but with different mechanism involved and different outcomes also affecting the system time response. The obtained results say that the traditional way to employ a multiple dye doping of the polymeric matrix, enables a scintillation yield increment due to the maximization of the system global emission efficiency. However, the system time response is worsened, because of the slow energy transfer steps involved. Conversely, in the system loaded with dense NPs the scintillation efficiency enhancement is due to the locally-sensitized generation of diffusing free charges, which enables their more effective recombination into emissive molecular excitons leaving substantially unchanged the scintillation time response. These results are particularly interesting because they demonstrate how the NPs presence has a huge effect on the system performance even at a very low concentration that preserves the optical and light transport properties of the material. The fine control of the composition at the nanoscale is therefore experimentally confirmed as a key tool to manage the properties of polymeric scintillators, in order to overcome the actual limitations. The obtained results open the way to additional explorations where also the effect of NPs size and electronic properties on the scintillation performance of composite systems can be investigated.

**ASSOCIATED CONTENT**
Supporting Information. The Supporting Information is available free of charge on the ACS Publications website at DOI: XXXXX.

Experimental methods, additional data and figures that support the structural and optical characterization of the conjugated molecules and scintillators, together with the details on the energy transfer modeling and the light/matter modeling employed.

**ORCID**
I. Villa 0000-0002-6150-7847
A. Monguzzi 0000-0001-9768-4573
R. Lorenzi 0000-0002-6199-0971
M. Orfano 0009-0002-9072-053X
V. Babin 0000-0003-3072-2242
F. Hájek 0000-0002-9344-4174
K. Kuldová 0000-0003-1053-0809
R. Kučerková 0000-0001-9441-0681
A. Beitlerová 0000-0001-7979-5544
I. Mattei 0000-0002-6819-1646
V. Čuba 0000-0002-6401-8117
L. Prouzová Procházková 0000-0002-2722-8094
M. Nikl 0000-0002-2378-208X

**Author Contributions**
The manuscript was written through contributions of all authors. All authors have given approval to the final version of the manuscript.

**Notes**
The authors declare no competing financial interest.

**ACKNOWLEDGMENTS**
I.V., A.M., and R.L. acknowledges financial support from the Italian Ministry of University (MUR) through grant no. PRIN 2020-SHERPA no. H45F2100343000, grant MINERVA LuMIminesceNt scintillating hEterostructures foR adVanced medical imaging no. H25E22000490006 and grant LUMINANCE LUMINescent gArNet CEramics and nanocrystals: material properties understanding and scintillation applications no. H53D23004540001. M.O. acknowledge support from the European Community through the grant no. 899293, HORIZON 2020 - SPARTE FET OPEN. I.V. and M. N. acknowledge support from the Marie Skłodowska-Curie Actions Widening Fellowships (MSCA-WF) grant no. 101003405-HANSOME. The work is also supported by OP JAC financed by ESIF and the MEYS SENDISO - CZ.02.01.01/00/22_008/0004596. Financial support from the Czech Science Foundation under Grant No. 23-05615S is gratefully acknowledged.

We thank you Dr. Melissa Saibene from the Piattaforma di Microscopia laboratory at the University of Milano-Bicocca for the TEM imaging on nanocomposites.

**TOC**

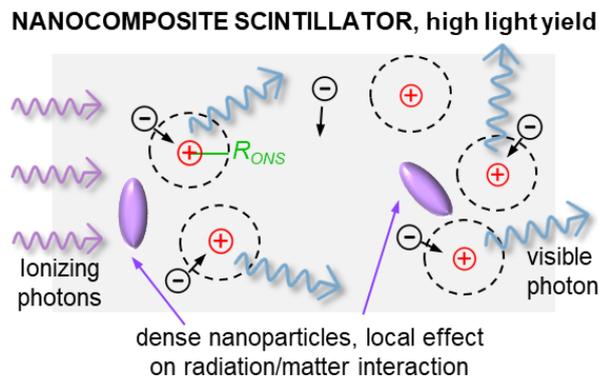